# A conditionally integrable Schrödinger potential of a bi-confluent Heun class

**T A Ishkhanyan[1,2,3], A M Manukyan[1,4] and A M Ishkhanyan[1,4]**

[1] Institute for Physical Research, NAS of Armenia, Ashtarak, 0203Armenia
[2] Moscow Institute of Physics and Technology, Dolgoprudny, 141700 Russia
[3] LICB, UMR CNRS 6303-Université de Bourgogne, Dijon, 21078 France
[4] Russian-Armenian University, H. Emin 123, Yerevan, 0051Armenia

**Abstract.** We present a bi-confluent Heun potential for the Schrödinger equation involving inverse fractional powers and a repulsive centrifugal-barrier term the strength of which is fixed to a constant. This is an infinite potential well defined on the positive half-axis. Each of the fundamental solutions for this conditionally integrable potential is written as an irreducible linear combination of two Hermite functions of a shifted and scaled argument. We present the general solution of the problem, derive the exact energy spectrum equation and construct a highly accurate approximation for the bound-state energy levels.

**1. Introduction**

We introduce a generalization of the recently presented exactly solvable inverse square root potential for the one-dimensional stationary Schrödinger equation [1]. This is a conditionally integrable singular potential which belongs to a bi-confluent Heun family of potentials involving negative half-integer powers of the space coordinate [2,3]. The term "conditionally integrable" is conventionally used to indicate that the parameters involved in the potential are not varied independently or a potential parameter is fixed to a constant. Well known examples of such potentials discussed by many authors in the past are the two Stillinger confluent hypergeometric potentials [4] and the Dutt-Khare-Varshni ordinary hypergeometric potential [5] (for other examples see, e.g., [6-12]).

The potential we introduce is given as

$$V = V_0 + \frac{5\hbar^2/(32m)}{x^2} + \frac{V_1}{x^{3/2}} - \frac{16m^2 V_1^3/\hbar^4}{\sqrt{x}}. \tag{1}$$

The shape of the potential is shown in figure 1. For $V_1 > 0$ this is a potential well defined on the positive half-axis $x > 0$ so that it sustains bound states. Because of the long-range component $\sim x^{-1/2}$ the number of the bound states is infinite. The potential is a member of the bi-confluent Heun family with $m_1 = -1$ (see [2,3]), which generalizes the second Stillinger [4] and the first Exton [7] potentials as well as the potential by López-Ortega [11].

We present the general solution of the one-dimensional Schrödinger equation for a particle of mass $m$ and energy $E$:

$$\frac{d^2\psi}{dx^2} + \frac{2m}{\hbar^2}(E - V(x))\psi = 0, \tag{2}$$

where $\hbar$ is the reduced Planck's constant, derive the exact equation for the energy spectrum and construct a highly accurate approximation for the bound-state energy levels.

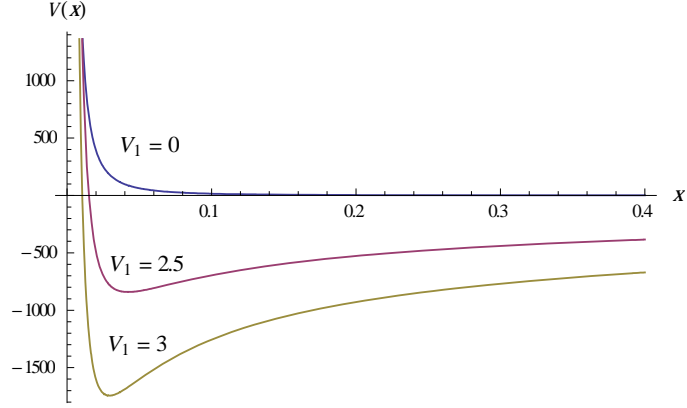

**Figure 1.** Potential (2) for $V_1 \geq 0$ ($m = \hbar = 1$, $V_0 = 0$).

A peculiarity of the solution is that none of the two independent fundamental solutions composing the general solution of the problem is written through the widely used one-term ansatz involving a single hypergeometric function. Rather, both fundamental solutions are written as irreducible linear combinations of two Hermite functions of a shifted and scaled argument.

We show that for bound states only one of the two fundamental solutions survives. This leads to an energy spectrum equation involving two Hermite functions of an energy-dependent auxiliary parameter. Applying an accurate approximation for these Hermite functions, we derive a highly accurate approximation for the energy levels.

## 2. The general solution

The solution of the Schrödinger equation for potential (1) is derived by reducing equation (2) to the bi-confluent Heun equation [13-15], which is one of the four confluent forms of the general Heun equation [16]. This is a second-order linear differential equation which has one regular singularity conventionally located at the origin and an irregular singularity of rank 2 conventionally located at the infinity. In an auxiliary five-parametric representation the bi-confluent Heun equation is written as

$$\frac{d^2 u}{dz^2} + \left(\frac{\gamma}{z} + \delta + \varepsilon z\right)\frac{du}{dz} + \frac{\alpha z - q}{z} u = 0. \tag{3}$$

This equation is widely encountered in various domains of current physics research such as quantum mechanics, quantum gravity, general relativity, solid state physics, etc. (see [13-15]). For example, in particle physics the equation appears when studying the quarkonium and related models describing the confinement of quarks and antiquarks [17,18].

The reduction of the Schrödinger equation to the bi-confluent Heun equation is presented in Refs. [2,3] (see also [19]). The next step is expanding the solution of the bi-confluent Heun equation (3) in terms of the Hermite functions as described in [3]. Then, by requiring the series to terminate on the third term and allowing the potential parameters to be dependent, we arrive at the following potential:

$$V = V_0 + \frac{5\hbar^2}{32mx^2} + \frac{V_1}{x^{3/2}} + \frac{V_2}{x} + \frac{8mV_1\left(-2mV_1^2 + \hbar^2 V_2\right)}{\hbar^4 \sqrt{x}}. \tag{4}$$

Choosing here $V_2 = 2mV_1^2/\hbar^2$ cancels the last term thus reproducing the López-Ortega potential [1] with an additional centrifugal-barrier term $5\hbar^2 x^{-2}/(32m)$, while by putting $V_2 = 0$ we cancel the coulombic term and get potential (1).

Omitting the details in order to avoid the text overlap (since the calculations are merely technical and exactly follow the lines of [3]), the resultant three-term Hermite-function solution of the bi-confluent Heun equation (3) for the given set of the potential parameters leads to the following general solution of the Schrödinger equation. The two independent fundamental solutions composing the general solution for real positive $V_1 > 0$ and $E < V_0$ are written as

$$\psi_F(x) = x^{-1/4} e^{-y^2/2} \left( H_{a+1/2}(y) + \left(\sqrt{2a} + A\sqrt[6]{2a}\right)\left(1 + \frac{\varepsilon \hbar^2 \sqrt{x}}{4mV_1}\right) H_{a-1/2}(y) \right), \tag{5}$$

where $H$ is the Hermite function of the shifted and scaled argument

$$y = \sqrt{-\varepsilon} x - \sqrt{2a} \tag{6}$$

and the involved parameters $a$ and $\varepsilon$ are given as

$$a = -\frac{2^{11} m^6 V_1^6}{\hbar^{12} \varepsilon^3}, \quad \varepsilon = \pm \sqrt{\frac{8m(-E + V_0)}{\hbar^2}}. \tag{7}$$

The parameter $A$ involved in the fundamental solutions (5) is equal to 1 if the minus sign for $\varepsilon$ is chosen in the second equation of (7) and $A = e^{4i\pi/3}$ for the plus sign of $\varepsilon$. The general solution of the Schrödinger equation is thus written as

$$\psi(x) = c_1 \psi_F^- \big|_{A=1} + c_2 \psi_F^+ \big|_{A=e^{4i\pi/3}}, \tag{8}$$

where $c_{1,2}$ are arbitrary constants. We note that this solution can alternatively be written in terms of the confluent hypergeometric functions.

## 3. Bound states and energy spectrum

The energy spectrum is derived by demanding the wave function to vanish at the origin and at infinity (see the discussion in [20]). An observation now is that the second fundamental solution diverges at infinity for any set of $V_1 > 0$ and energy $E < V_0$. This is readily understood if one recalls that the Hermite function behaves at infinity as a power of its argument: $H_\nu(y) \sim 2^\nu y^\nu$, so that the asymptote of the fundamental solution (5) at infinity is governed by the pre-factor $e^{-y^2/2}$. Since $-y^2 \sim \varepsilon x$, it is understood that the second fundamental solution $\psi_F^+$, for which $\varepsilon > 0$, exponentially diverges, while the first fundamental solution $\psi_F^-$ with $\varepsilon < 0$ always vanishes at $x \to +\infty$. Hence, the requirement for the wave function to vanish at infinity is satisfied only if $c_2 = 0$. The requirement for the wave function to vanish at the origin then immediately leads to the following equation:

$$H_{a+1/2}\left(-\sqrt{2a}\right) + \left(\sqrt{2a} + \sqrt[6]{2a}\right) H_{a-1/2}\left(-\sqrt{2a}\right) = 0. \tag{9}$$

This is the exact equation for the energy spectrum, which is constructed through relations (7). The equation possesses a countable infinite set of positive roots $a_n$. Except the smallest root $a_0 = 1/2$, each of these roots stands for a bound state energy. The root $a_0 = 1/2$ should be disregarded because for this $a$ the wave function becomes zero. The next root, which is close to $3/2$, produces the first bound state with a single maximum. We prescribe to this root the number 1.

An accurate approximation for the spectrum equation (9) is achieved by noting that the arguments and indexes of the involved Hermite functions $H_\nu(z)$ belong to the left transient layer for which $z \approx -\sqrt{2\nu}$ [21]. We then apply the Airy-function approximation (which is further reduced to trigonometric functions) for the Hermite functions suggested in [22] to derive the approximation

$$F \equiv 1 + \left(\sqrt{2a} + \sqrt[6]{2a}\right) \frac{H_{a-1/2}\left(-\sqrt{2a}\right)}{H_{a+1/2}\left(-\sqrt{2a}\right)} \approx \frac{\sin(\pi a - \pi/3) - 6B_0 2^{-1/3} \sin(\pi a + \pi/3)}{\sin(\pi a - \pi/3) + 6B_0 a^{1/3} \sin(\pi a + \pi/3)}, \tag{10}$$

where

$$B_0 = \frac{\Gamma(1/3)}{6\sqrt[3]{3}\,\Gamma(2/3)}. \tag{11}$$

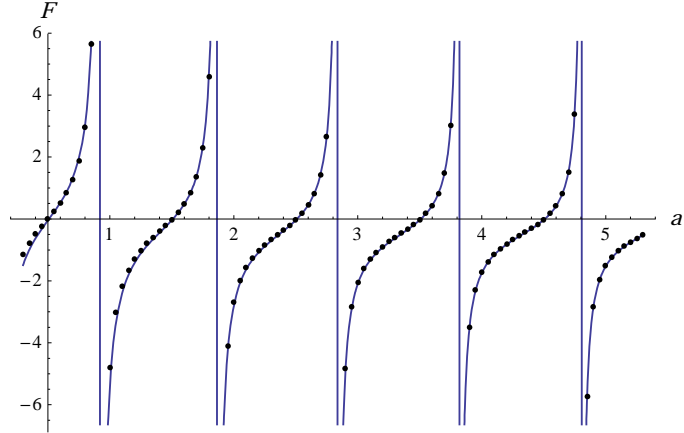

**Figure 2.** Filled circles – exact numerical result, solid curves – approximation (10).

The derived approximation is compared with the exact function $F$ in figure 2. As seen, we have a highly accurate result. Hence, as it follows from equation (9), the spectrum equation is well approximated by the simple trigonometric equation

$$\sin(\pi a - \pi/3) - 6B_0 2^{-1/3}\sin(\pi a + \pi/3) = 0. \qquad (12)$$

It is immediately seen from figure 2 that the roots of this equation (and thus the roots of the exact spectrum equation (9)) are close to positive half-integers. This result is readily deduced if we note that $6B_0 2^{-1/3} \approx 1$ so that the approximate spectrum equation is finally reduced to the very simple equation $-\sqrt{3}\cos(a\pi) = 0$. Thus, $a_n = n + 1/2$, $n \in \mathbb{N}$. According to equation (7), the energy spectrum is then given by the compact formula

$$E_n \approx -\frac{32 m^3 V_1^4 / \hbar^6}{(2n+1)^{2/3}}, \quad n = 1, 2, 3, \ldots. \qquad (13)$$

It is understood that this result stands for the semiclassical limit for large $n \to \infty$, the Maslov correction index [22,23] being 1/2. However, this is a fairly good approximation for all levels. The relative error is less than $2 \times 10^{-3}$ even for the ground state (see the inset of figure 3). Having this result, it is not difficult to calculate the next approximations, e.g., by expanding equation (10) in terms of powers of $(2n+1)^{-2/3}$. The second approximation then provides relative error less than $10^{-5}$ for all energy levels. The first three non-normalized wave functions are shown in figure 4.

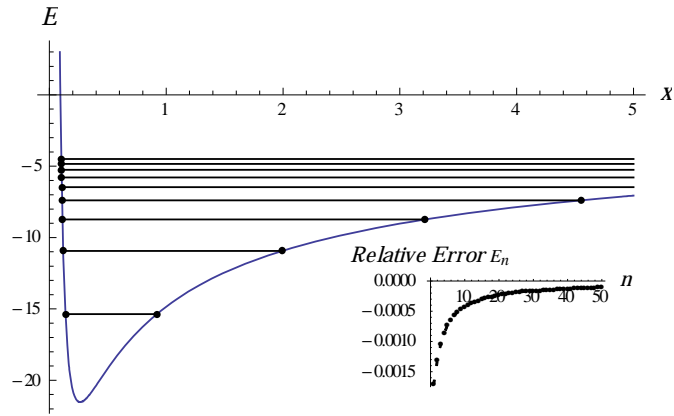

**Figure 3.** Energy spectrum of potential (1) for $V_1 = 1$ ($m = \hbar = 1$, $V_0 = 0$).

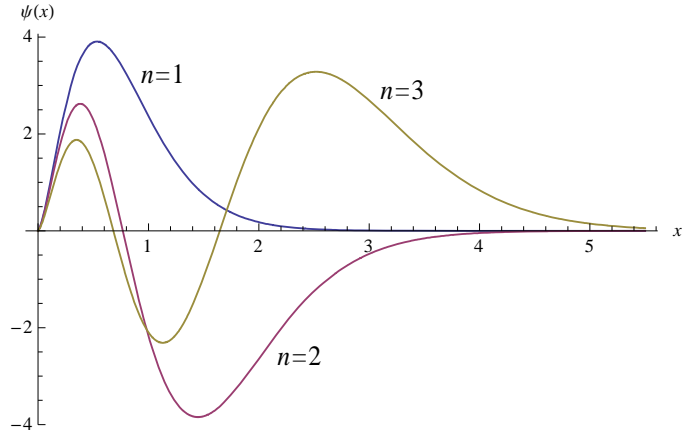

**Figure 4.** The first three non-normalized wave functions for $V_1 = 1$ ($m = \hbar = 1$, $V_0 = 0$).

## 4. Discussion

Discussing the reducibility of the Schrödinger equation to the bi-confluent Heun equation, Lemieux and Bose have presented five families of potentials [2] of which four involve power-law terms and the fifth one is a generalization of the Morse potential which involves exponentials [24]. These are rich families including, as particular cases, the three exactly solvable classical confluent hypergeometric potentials (harmonic oscillator, Coulomb and Morse) and many of the known conditionally integrable potentials discussed in the past.

The potential we have introduced belongs to the first Lemieux-Bose family which involves negative integer and half-integer power-law terms. This is a remarkable family the known members of which include, for instance, the second Stillinger [4] and the first Exton [7] potentials as well as the inverse square root potential [1], the potential by López-Ortega [11] and the recent generalization of the latter two potentials [12].

Because of its wide appearance, the solutions of the bi-confluent Heun equation have been studied by many authors [13-15]. Among the recent developments, of particular interest for this research are the expansions of the solutions in terms of special functions other than simple powers [25,26], in particular, the expansion in terms of the Hermite functions of a shifted and scaled argument [3]. The latter expansion indicates that there exists a large set (presumably, infinite) of exactly or conditionally integrable potentials the solution for which is written as a linear combination with constant coefficients of a finite number of Hermite functions. The first potentials of this set are the classical harmonic oscillator potential and the two potentials by Stillinger [4] the solution for which involves a single Hermite function. The next come the potentials for which the fundamental solutions present irreducible combinations of two Hermite functions. These are the two Exton potentials [7] the first of which involves the inverse square root potential [1] and its conditionally integrable generalization [12] which includes the López-Ortega supersymmetric pair of potentials [11].

The potential we have presented is just the next in the list, i.e., the solution of the Schrödinger equation for this potential is a linear combination with constant coefficients of three Hermite functions. Because these are contiguous functions, using the known recurrence relations, the combination is reduced to a combination with non-constant coefficients of two Hermite functions. We have presented the general solution of the Schrödinger equation for this potential and have derived the exact energy spectrum equation for the bound states. Using an accurate approximation for the involved Hermite functions, we have approximated this equation by a simple transcendental equation involving trigonometric functions. We have seen that the solution of the latter equation provides a highly accurate approximation for all bound-state energy levels: the relative error is less than $2 \times 10^{-4}$ even for the ground state.

**Acknowledgments**
This research has been conducted within the scope of the International Associated Laboratory IRMAS (CNRS-France & SCS-Armenia). The work has been supported by the Armenian State Committee of Science (Grant No. 15T-1C323) and the Armenian National Science and Education Fund (ANSEF Grant No. PS-4558). T.A. Ishkhanyan acknowledges the support from SPIE through a 2017 Optics and Photonics Education Scholarship and thanks the French Embassy in Armenia for a doctoral grant.